\documentclass[a4]{epl2}
\usepackage{graphicx}
\usepackage{epsfig}
\usepackage{color}
\vspace{-2cm}
\title{\textcolor[rgb]{0.00,0.00,0.00}{Maxwellian quantum mechanics}}
\shorttitle{\textcolor[rgb]{0.54,0.17,0.89}{Maxwellian quantum mechanics}}

\author{A. I. Arbab\inst{}\footnote{arbab.ibrahim@gmail.com}}

\institute{
Department of Physics, College of Science, Qassim University, P.O. Box 6644, 51452 Buraidah,  KSA}


\abstract{Expanding the ordinary Dirac's equation in quaternionic form  yields Maxwell-like field equations. As in
the Maxwell's formulation, the particle fields are represented by a scalar, $\psi_0$ and a vector $\vec{\psi}$. The analogy with Maxwell's equations requires that the inertial fields are $\vec{E}_D=c^2\vec{\alpha}\times\vec{\psi}$, and
$\vec{B}_D=\vec{\alpha}\,\psi_0+c\beta\,\vec{\psi}$ and that $\psi_0=-c\beta\,\vec{\alpha}\cdot\vec{\psi}$, where $\beta$, $\vec{\alpha}$ and $c$ are the Dirac matrices and the speed of light, respectively. An alternative solution suggests that magnetic monopole-like behavior accompanies Dirac's field. In this formulation, a field-like representation of Dirac's particle is derived. It is shown that when the vector field of the particle, $\vec{\psi}$, is
normal to the vector $\vec{\alpha}$, Dirac's field represents a medium with maximal conductivity. The energy flux (Poynting  vector) of the Dirac's fields is found to flow in opposite direction to the particle's motion.  A system of equivalently symmetrized Maxwell's equations is introduced. A longitudinal (scalar) wave traveling at speed of light is found to  accompany magnetic charges flow. This wave is not affected by presence of electric charges and currents. The Lorentz boost transformations of the matter fields are equivalent to $c\vec{\psi}\,' =c\vec{\psi}\pm\beta\vec{\alpha}\,\psi_0\,,\psi_0\,'=\psi_0\mp c\beta\vec{\alpha}\cdot\vec{\psi}\,.$}

\begin{document}
\vspace{-2cm}
\maketitle
\baselineskip=20pt

\section{\textcolor[rgb]{0.00,0.07,1.00}{Introduction}}
Maxwell had unified the laws of electricity and magnetism in a
consistent way into a set of four equations \textcolor[rgb]{0.00,0.07,1.00}{\cite{griff}}. The sources of the electric
and magnetic fields are the charges and currents (moving charges). De
Broglie had postulated that a moving particle exhibits a wave-like
nature. This wave is concomitant with the particle as it moves on in
a form of a wave packet centered around the particle. On the other
hand, the electric and magnetic fields associated with the moving charges and currents
spread away in space and have oscillatory behavior. Biot-Savart
showed that the electric and magnetic fields due to a charged
particle moving at constant speed are stationary and centered around
the particle. In quantum mechanics, Dirac and Schrodinger equations
 govern the motion of a particle (e.g., an electron) owing to its mass (matter)
nature irrespective of its charge. However, a moving charged particle  can interact with the electromagnetic field existing in
space. The proper formation is described by electrodynamics merging
quantum mechanics with Maxwell's theory.

The question that normally
arises is that how do the two waves interact, the matter waves (due
to mass) and the field waves (due to charge). Moreover, one can also
think of another wave nature, that is  governed by some equation, pertaining to another physical property of the particle, like spin, instead
of incorporating it in the former formulations.

In a recent paper, we have derived the ordinary Dirac equation (in an unfamiliar form), Klein-Gordon, and Schrodinger equations from a quaternionic form of Dirac's equation \textcolor[rgb]{0.00,0.07,1.00}{\cite{arb1}}. Earlier, we have made an analogy between hydrodynamics and electrodynamics and showed that hydrodynamics laws can be written in  Maxwell-like fields equations. Following a same line of reasoning, we have presented an analogy between matter (de Broglie) waves and the electromagnetic waves \textcolor[rgb]{0.00,0.07,1.00}{\cite{arb2}}. Moreover, a quaternionic Maxwell's equation is found to allow for scalar (longitudinal) waves besides the ordinary electromagnetic transverse waves \textcolor[rgb]{0.00,0.07,1.00}{\cite{arb3}}. The use of quaternions in formulating physical laws is found to be very rich \textcolor[rgb]{0.00,0.07,1.00}{\cite{Arb, arb}}.

In this work we extend our formulation to investigate another form of quaternionic Dirac's equation. \textcolor[rgb]{1.00,0.00,0.00}{We aim in this work to describe the Dirac particle (matter nature) by assigning a field
character rather than a matter-wave character. We then compare the equations governing these  fields with the corresponding Maxwell's equations. This approach will enhance deepening the notion of the duality
nature exhibited by micro-particles. However, a rather complete wave-particle duality is recently investigated \textcolor[rgb]{0.00,0.07,1.00}{\cite{arb4}}.  The new quaternionic Dirac's equation formulation is found to be very promising
in bringing the above ideas into reality. The resulting set of equations emerging from the quaternionic Dirac's equation are analogous to Maxwell's electromagnetic field equations.}

We introduce the classical electromagnetism in Section 1, and the quaternionic Dirac's equation in Section 2. We then define the analogous electric and magnetic matter fields, and compare the latter field with that of the Biot-Savart fields arising from a charge moving at constant velocity. In Section 3 we provide the energy and momentum densities of the two analogous fields, and showed that they are described by the same set of equations. We have found that while the electromagnetic fields can dissipate energy in some cases, the matter fields spread without energy loss thus preserving the particle integrity. The direction of the matter field flux density is opposite to the particle's motion. A symmetrised Dirac's field equations are obtained by relaxing one of our conditions on the Dirac's fields. This latter condition allows for longitudinal wave nature for the Dirac's field and permits magnetic monopoles to exist. This  longitudinal (scalar) wave, $\Lambda$, has an energy flux $-\Lambda\,\vec{B}$ and energy density $\frac{\varepsilon_0}{2}\,\Lambda^2$.

\section{\textcolor[rgb]{0.00,0.07,1.00}{Maxwell's fields equations}}
The force on a charged particle $q$ in the presence of electric
$\vec{E}$, and magnetic field $\vec{B}$, is given by Lorentz force
as
\begin{equation}
\vec{F}_L=q(\vec{E}+\vec{v}\times\vec{B})\,.
\end{equation}
According to the Newton's second law of motion the particle (charge)
acceleration $\vec{a}$ is given by $\vec{F}_L=m\vec{a}$. However,
the electromagnetic fields are determined by Maxwell's equations
\textcolor[rgb]{0.00,0.07,1.00}{\cite{griff}},
\begin{equation}\label{1}
\vec{\nabla}\times\vec{E}=-\frac{\partial \vec{B}}{\partial t}\,,
\end{equation}
\begin{equation}
\vec{\nabla}\times\vec{B}=\mu_0\vec{J}+\frac{1}{c^2}\frac{\partial
\vec{E}}{\partial t}\,,
\end{equation}
and
\begin{equation}\label{1}
\vec{\nabla}\cdot\vec{E}=\frac{\rho}{\varepsilon_0}\,,
\end{equation}
\begin{equation}
 \vec{\nabla}\cdot\vec{B}=0\,.
\end{equation}
Here $\vec{J}$ and $\rho$ represent the source of the
electromagnetic fields. In Maxwell's formulation the matter field
hasn't been considered, as this was not yet known at that time.

We would like now to employ quaternions to rewrite Dirac's equation,
and to compare the resulting equations with Maxwell's equations.
This is because the electron charge nature is expressed by Maxwell's
equations, while its wave nature is described by Dirac. To this aim,
we should express Dirac particles by matter fields rather than
wavefunctions (spinors).

\section{\textcolor[rgb]{0.00,0.07,1.00}{The quaternionic Dirac's equation}}

The ordinary Dirac's equation of a spin-1/2 particle with rest mass $m$ is expressed as \textcolor[rgb]{0.00,0.07,1.00}{\cite{drell}}
\begin{equation}\label{1}
p^\mu\gamma_\mu \psi=mc\,\psi\,,
\end{equation}
where $\gamma^\mu$ are expressed in terms of Pauli matrices, $c$ is the speed of light, and
$\psi$ are the spinors representing the Dirac's wavefunction. In
quaternionic form Equation  (6) reduces to
\begin{equation}\label{1}
\tilde{P}\tilde{\gamma} \,\tilde{\Psi}=mc\,\tilde{\Psi}\,,
\end{equation}
where
\begin{equation}\label{1}
\tilde{P}=\left(\frac{i}{c}\,E\,, \vec{p}\right)\,,\qquad
\tilde{\gamma}=(i\beta\,, \vec{\alpha})\,,\qquad
\tilde{\Psi}=\left(\frac{i}{c}\,\psi_0\,, \vec{\psi}\right)\,.
\end{equation}
In quantum mechanics the above energy and momentum become operators, \emph{viz.}, $\vec{p}=-i\hbar\vec{\nabla}$ and $E=i\hbar\,\frac{\partial}{\partial t}$. Here $\psi_0$ and $\vec{\psi}$ are the  Dirac's
particle fields which we will shortly give them a meaning. To envisage this relation, we know that the Dirac spinor can be decomposed into $2\times 2$ components (\emph{e.g.},  Weyl). From group point of view, one has $2\bigotimes 2=1\bigoplus 3$. Here $\vec{\psi}$ represents the 3-vector and $\psi_0$  the 1-scalar. Hence, the 4-components spinor is decomposed in a scalar and a vector. Some authors express Maxwell's equations in a Dirac-like form \textcolor[rgb]{0.00,0.07,1.00}{\cite{vector1, vector2}}. They employ a complex vector, $\vec{F}=\frac{\vec{E}}{c}+i\vec{B}$. This vector acts as a wavefunction of the photon \textcolor[rgb]{0.00,0.07,1.00}{\cite{vector3}}. The analogy they drew between the new Maxwell form and Dirac equation was not complete. That  is because Dirac spinors have 4-components, while the complex vector has 3-components. In our present quaternionic representation, we have a scalar function besides the 3-vector, $\vec{F}$. Note that the original Maxwell's equations admit only transverse wave, but not  scalar wave. Our present formulation allows for scalar wave besides the transverse wave. Hence, the 3-vector, $\vec{\psi}$ and a 1-scalar, $\psi_0$ can be related to the 3-vector, $\vec{F}$ and additional scalar electromagnetic scalar ($\Lambda$). Consequently, a scalar wave is admissible in the quaternionic Maxwell-Dirac formulation.

Let us now employ the quaternion multiplication
rule for two quaternions, $\tilde{A}=(a_0\,, \vec{a})$ and
$\tilde{B}=(b_0\,, \vec{b})$, \emph{viz.} \textcolor[rgb]{0.00,0.07,1.00}{\cite{arb3}}
\begin{equation}\label{1}
\tilde{A}\,\tilde{B}=(a_0\,b_0-\vec{a}\cdot\vec{b}\,\,,
\,\,a_0\,\vec{b}+\vec{a}\,b_0+\vec{a}\times\vec{b})\,.
\end{equation}
Now apply Equations (8) and (9) in Equation  (7) and then equate the real and
imaginary parts to each other, to obtain
\begin{equation}\label{1}
\vec{\nabla}\cdot(c^2\vec{\alpha}\times\vec{\psi})=\frac{mc^2}{\hbar}\,\psi_0\,,
\end{equation}
\begin{equation}\label{1}
\vec{\nabla}\times(\vec{c^2\alpha}\times\vec{\psi})=-\frac{\partial}{\partial
t}(\vec{\alpha}\,\psi_0+c\beta\,\vec{\psi})+\vec{\nabla}(c^2\vec{\alpha}\cdot\vec{\psi}+c\beta\,\psi_0)\,,
\end{equation}

\begin{equation}\label{1}
\vec{\nabla}\times(\vec{\alpha}\,\psi_0+c\beta\,\vec{\psi})-\frac{1}{c^2}\frac{\partial}{\partial
t}\,(c^2\vec{\alpha}\times\vec{\psi})=\frac{mc^2}{\hbar}\,\vec{\psi}\,,
\end{equation}
and
\begin{equation}\label{1}
-\vec{\nabla}\cdot(c\,\vec{\psi}+\beta\,\vec{\alpha}\,\psi_0)+\frac{\partial}{\partial
t}\,\left(\frac{\psi_0}{c}+\beta\,\vec{\alpha}\cdot\vec{\psi}\right)=0\,.
\end{equation}
To reproduce a set of equations similar to Maxwell's equations we
make the following choices:
\begin{equation}\label{1}
\vec{E}_D=c^2\vec{\alpha}\times\vec{\psi}\,,
\end{equation}
\begin{equation}
\vec{B}_D=\vec{\alpha}\,\psi_0+c\beta\,\vec{\psi}\,,
\end{equation}
\begin{equation}
\vec{J}_D=\frac{m\,c^2}{\mu_0\hbar}\,\,\vec{\psi}\,,
\end{equation}
\begin{equation}
\rho_D=\frac{m}{\mu_0\hbar}\,\psi_0\,,
\end{equation}
provided that
\begin{equation}
\psi_{0}=-\,c\beta\,\vec{\alpha}\cdot\vec{\psi}\,.
\end{equation}
This particular plausible  linear combinations of Dirac vector and scalar, reproduces Maxwell's-like equations. This implies that the quaternionic Dirac equation is very rich. Maxwell's equations are known to describe spin -1 particle (photon) which is represented by a 3 - vector, while Dirac's equation describes spin - $\frac{1}{2}$ particles which are represented by 4-component spinors. In our quaternionic Dirac's equation, our particle is described by 4 - components (a scalar and a 3-vector). Hence, it formally describes a spin - 0 (scalar) and spin - 1 (vector) particles. Note that the group decomposition of two spin-$\frac{1}{2}$ particles is equal to a scalar and 3-vector, \emph{i.e.}, $2\bigotimes2=1\bigoplus3$.
Since $\vec{\alpha}$ and $\beta$ are  $\times4$ matrices, then $\vec{E}_D$ and $\vec{B}_D$ have 4 - components. The 4 - $\vec{E}_D$ components are that of static fields arising from a particle and its antiparticle. The 4 - $\vec{B}_D$ components are that of static fields arising from a magnetic charge and its antimagnetic charge. Hence, Dirac's equation inherently encompasses magnetic charges. This is in agreement with the Dirac's quantization rule for magnetic charge \textcolor[rgb]{0.00,0.07,1.00}{\cite{dirac}}.

Applying Equations (14) - (18) in Equations (10) - (13) yields
\begin{equation}\label{1}
\vec{\nabla}\cdot(c^2\vec{\alpha}\times\vec{\psi})=\frac{mc^2}{\hbar}\,\psi_0\,,
\end{equation}
\begin{equation}\label{1}
\vec{\nabla}\times(c^2\vec{\alpha}\times\vec{\psi})=-\frac{\partial}{\partial
t}\,(\vec{\alpha}\,\psi_0+c\beta\,\vec{\psi})\,,
\end{equation}

\begin{equation}\label{1}
\vec{\nabla}\times(\vec{\alpha}\,\psi_0+c\beta\,\vec{\psi})=\frac{mc^2}{\hbar}\,\vec{\psi}+\frac{1}{c^2}\frac{\partial}{\partial
t}\,(c^2\vec{\alpha}\times\vec{\psi})\,,
\end{equation}
and
\begin{equation}\label{1}
\vec{\nabla}\cdot(\vec{\alpha}\,\psi_0+c\,\beta\,\vec{\psi})=0\,.
\end{equation}
Applying Equations (14) - (18) in Equations (19) - (22) yields the
Dirac-Maxwell's-like equations
\begin{equation}\label{1}
\vec{\nabla}\cdot\vec{E}_D=\frac{\rho_D}{\varepsilon_0}\,,
\end{equation}
\begin{equation}\label{1}
\vec{\nabla}\times\vec{E}_D=-\frac{\partial \vec{B}_D}{\partial
t}\,,
\end{equation}
\begin{equation}
\vec{\nabla}\times\vec{B}_D=\mu_0\vec{J}_D+\frac{1}{c^2}\frac{\partial
\vec{E}_D}{\partial t}\,,
\end{equation}
and
\begin{equation}
 \vec{\nabla}\cdot\vec{B}_D=0\,.
\end{equation}
According to the aforemention ansatzs, we expect the equation of motion of the particle to be governed by Lorentz-like force, \emph{viz.},
$$\vec{F}_D=m(\vec{E}_D+\vec{v}\times\vec{B}_D)\,.$$
Applying Equations (14) and (15) yields
$$\vec{F}_D=0\,,\qquad {\rm if }\qquad \vec{v}=c\beta\vec{\alpha}\,.$$
It is interesting to see that the inertial electric and magnetic forces are equal and opposite.
This implies that the particle acceleration  is zero, as expected.
It is remarkable to see that Equations (16) and (17) can be expressed as
\begin{equation}
\rho_D=\frac{1}{2}\,\sigma_m\,\psi_0\,,\qquad
\vec{J}_D=\frac{1}{2}\,\sigma_m\,c^2\,\vec{\psi}\,,\qquad
\sigma_m=\frac{2m}{\mu_0\hbar}
\end{equation}
Here $\sigma_m$ represents the maximal conductivity that any system can attain as recently hypothesized \textcolor[rgb]{0.00,0.07,1.00}{\cite{arb5, arb2}}. Hence, Equations (14) and (27) yield $|\vec{E}_D|=c^2|\vec{\alpha}|\,|\vec{\psi}|$ and
$|\vec{J}_D|=\sigma_mc^2|\vec{\psi}|$, where $|\vec{\alpha}|=1$. Thus, the Ohm's law, $\vec{J}=\sigma\, \vec{E}$ is satisfied for a medium with maximal conductivity where $\vec{\alpha}$ is normal to
$\vec{\psi}$. This shows that Equation  (24) is applicable to a conducting
medium  that has a maximal conductivity, \emph{i.e.},
$\sigma=\sigma_m$. Such systems (media) are recently shown to exist
in white dwarfs and neutron stars \textcolor[rgb]{0.00,0.07,1.00}{\cite{arb6}}. Hence, the Dirac's
field is equivalent to  Maxwell fields propagating in a medium with
maximal conductivity. In Dirac's theory $\vec{\alpha}$ is related to
the spin of the particle \textcolor[rgb]{0.00,0.07,1.00}{\cite{drell}}.

Since the electric and magnetic fields have wave character, the
Dirac's fields will have a wave character too. Note that as apparent from Equations (23) - (26) together with Equations (16) and (17) that the fields of a massless Dirac's particle propagate with speed of light.

The system of equations, Equations (10)-(13) or Equations (19)-(22), satisfies
the continuity equation, \emph{i.e.},
\begin{equation}
\vec{\nabla}\cdot\vec{J}_D+\frac{\partial\rho_D}{\partial t}=0\,.
\end{equation}
This is evident if we take the divergence of Equation  (21) and use Equation  (19)
to obtain
\begin{equation}
\vec{\nabla}\cdot\left(\frac{mc^2}{\mu_0\hbar}\right)\vec{\psi}+\frac{\partial}{\partial
t}\left(\frac{m\psi_0}{\mu_0\hbar}\right)=0\,.
\end{equation}
This is in agreement with the definition in Equations (16) and (17). Note
that in electromagnetism the electric and magnetic fields in the
rest frame  of a charge moving with constant velocity satisfy \textcolor[rgb]{0.00,0.07,1.00}{\cite{griff}}
\begin{equation}
\vec{v}\cdot\vec{E}=\vec{v}\cdot\vec{B}=0\,,\qquad \vec{E}=
-\vec{v}\times\vec{B}\,,\qquad
\vec{B}=\frac{\vec{v}\times\vec{E}}{c^2}\,.
\end{equation}
Equation (30) describes  the fields of a point charge moving at
constant velocity. These equations are usually valid for $v<<c$. These fields
are associated with the particle motion and aren't radiated away. These are
often expressed by Biot-Savart law. We obtain from Equations (15) and (16) the
following equations
\begin{equation}
c\,\vec{\alpha}\cdot\vec{E}_D=c\,\vec{\alpha}\cdot\vec{B}_D=0\,,\qquad
\vec{E}_D= -\beta\,c\vec{\alpha}\times\vec{B}_D\,,\qquad
\vec{B}_D=\frac{\beta\,c\vec{\alpha}\times\vec{E}_D}{c^2}\,.
\end{equation}
Note that in the Dirac's theory, $\vec{v}=c\,\vec{\alpha}$, but here we have $\vec{v}=c\,\beta\,\vec{\alpha}$. By
applying this definition in Equation  (31) and comparing the resulting equations with
Equation  (30), we disclose that the Dirac's fields are consistent with
Maxwell's fields. Moreover, we reveal that, since $\beta$ has
eigenvalues $\pm\,1$, the magnetic and electric fields of a moving
charged particle spread always in two directions (to the left and
the right of the particle motion) irrespective of the particle's
velocity direction.  That means,
$\vec{B}_D=\mp\,\frac{\vec{v}\times\vec{E}_D}{c^2}$ and
$\vec{E}_D=\pm\,\vec{v}\times\vec{B}_D$. The Lorentz force on the particle vanishes, as evident in substituting Equations (14) and (15) in (1). This is so because the particle is moving at constant velocity.

\subsection{\textcolor[rgb]{0.50,0.00,0.50}{Lorentz boosting transformations}}

Let us now consider the following transformation of the two fields, $\vec{\psi}$ and $\psi_0$ as
$$
\hspace{4cm}c\vec{\psi}\,' =c\vec{\psi}-\beta\vec{\alpha}\,\psi_0\,,\qquad\qquad \psi_0\,'=\psi_0+c\beta\vec{\alpha}\cdot\vec{\psi}\,,\hspace{1.5cm} (E1)
$$
Applying Equation  (E1) in Equations (14) - (17) yields
$$
\hspace{4cm}\vec{E}_D\,' =\vec{E}_D\,,\qquad \qquad\qquad \vec{B}_D\,' =-\frac{\vec{v}\times\vec{E}_D}{c^2}\,,\hspace{2.2cm} (E2)
$$
and
$$
\hspace{4cm}\vec{J}_D\,'=\vec{J}_D-\vec{v}\rho_D\,,\qquad\qquad \rho_D\,'= \rho_D+\frac{\vec{v}\cdot\vec{J}_D}{c^2}\,,\hspace{1.5cm} (E3)
$$
where we have used $\vec{v}=c\beta\,\vec{\alpha}$, and the prime quantity represents the quantity in the boosted frame. Apparently, the quaternion fields transform as 4-vectors under Lorentz boost. Notice that the mass and current densities transform in the same way as that of the quaternion fields. These are of the same nature of Lorentz boosting of the electromagnetic fields with small velocity. Hence, Equation  (E1) represents the Lorentz boosting of the matter field with a small velocity. Equation (E2) gives the  transformations of the electric matter field parallel to the direction of motion and a magnetic matter field perpendicular to the direction of motion. Other components are zeros.  However, if we had used the transformations
$$
\hspace{4cm}c\vec{\psi}\,' =c\vec{\psi}+\beta\vec{\alpha}\,\psi_0\,,\qquad\qquad \psi_0\,'=\psi_0-c\beta\vec{\alpha}\cdot\vec{\psi}\,.\hspace{1.5cm} (E4)
$$
we would have obtained
$$
\hspace{4cm}\vec{E}_D\,' =\vec{E}_D\,,\qquad \qquad\qquad \vec{B}_D\,' =2\vec{B}_D+\frac{\vec{v}\times\vec{E}_D}{c^2}\,,\hspace{1.3cm} (E5)
$$
and
$$
\hspace{4cm}\vec{J}_D\,'=\vec{J}_D+\vec{v}\rho_D\,,\qquad\qquad \rho_D\,'= \rho_D-\frac{\vec{v}\cdot\vec{J}_D}{c^2}\,,\hspace{1.5cm} (E6)
$$
Thus, in a frame where the new magnetic matter field vanishes (\emph{i.e.}, $\vec{B}_D\,'=0$ and hence $\vec{B}_D=-\frac{1}{2}\frac{\vec{v}\times\vec{E}_D}{c^2}$ ), the magnetic field in the rest frame picks up automatically the factor of 1/2 (Thomas factor) that was shown later, for the electromagnetic fields, to be due to  relativistic effect.

\section{\textcolor[rgb]{0.00,0.07,1.00}{The Dirac's fields energy}}

Owing to the analogy existing between Dirac's fields and Maxwell's
field, we can now find the energy and momentum densities of these
fields. The energy equation in Maxwell's theory is defined by \textcolor[rgb]{0.00,0.07,1.00}{\cite{griff}}
\begin{equation}
\frac{\partial u}{\partial
t}+\vec{\nabla}\cdot\vec{S}=-\vec{J}\cdot\vec{E}\,,\qquad
u=\frac{\varepsilon_0}{2}\,E^2+\frac{B^2}{2\mu_0}\,,\qquad
\vec{S}=\frac{\vec{E}\times\vec{B}}{\mu_0}\,.
\end{equation}
For Dirac's fields, the dissipation energy term,
$\vec{J}_D\cdot\vec{E}_D=\frac{mc^4}{\mu_0\hbar}\,\vec{\psi}\cdot(\vec{\alpha}\times\vec{\psi})=0$,
vanishes. Consequently the Dirac's fields don't lose energy when
spread out. Hence, the energy conservation equation for Dirac's
fields is given by
\begin{equation}
\frac{\partial u_D}{\partial t}+\vec{\nabla}\cdot\vec{S}_D=0\,,
\end{equation}
where $u_D$ and $\vec{S}_D$ are the corresponding energy and
momentum densities of the Dirac's fields. Using Equations (14) and (15)
together with Equations (32) and (18), one finds
\begin{equation}
\vec{S}_D=\frac{c^2\psi_0}{\mu_0}\,\vec{\psi}\,,\qquad
u_D=\frac{c^2}{\mu_0}\,\psi^2\,.
\end{equation}
This implies that the energy spreads out along the  vector field
($\vec{\psi}$). Since the magnetic field of a point charge is not
distributed uniformly, the Dirac's fields follow a similar pattern
too.

We now notice that Equations (16) and (17) suggest that
\begin{equation}
\vec{J}_D=\rho_D\, \vec{v}_D\,,\qquad
\vec{v}_D=\frac{c^2}{\psi_0}\,\vec{\psi}\,.
\end{equation}
This clearly shows that the Dirac's vector field of the particle
ushers in the velocity direction. Recall that the relativistic velocity of a particle is related to total energy $(E$) and momentum ($\vec{p}$) by a similar relation, \emph{viz}., $\vec{v}=\frac{c^2}{E}\,\vec{p}$. Moreover, Equation  (34) and (35) using Equation  (18) reveal
that
\begin{equation}
\vec{S}_D=-u_D\vec{v}_D\,.
\end{equation}
Hence, the energy flux outflows in the opposite direction of the
particle motion. This behaviour agrees with our earlier findings for matter waves in general \textcolor[rgb]{0.00,0.07,1.00}{\cite{arb2, long}}.

\section{\textcolor[rgb]{0.00,0.07,1.00}{New solution}}

Let us now assume that Equation  (18) is not satisfied, and instead let
\begin{equation}
\Lambda_D=c\beta\,\psi_0+c^2\vec{\alpha}\cdot\vec{\psi}\,.
\end{equation}
Substituting Equation  (37) in Equations (10) - (13) yields
\begin{equation}\label{1}
\vec{\nabla}\cdot\vec{E}_D=\frac{\rho_D}{\varepsilon_0}\,,
\end{equation}
\begin{equation}\label{1}
\vec{\nabla}\times\vec{E}_D=-\frac{\partial \vec{B}_D}{\partial
t}-\vec{J}_m\,,
\end{equation}
\begin{equation}
\vec{\nabla}\times\vec{B}_D=\mu_0\vec{J}_D+\frac{1}{c^2}\frac{\partial
\vec{E}_D}{\partial t}\,,
\end{equation}
and
\begin{equation}
 \vec{\nabla}\cdot\vec{B}_D=\rho_m\,,
\end{equation}
where $\vec{J}_m=-\vec{\nabla}\Lambda_D$, and $\rho_m=\frac{1}{c^2}\frac{\partial\Lambda_D}{\partial t}$. These are the magnetic current and mass densities associated with the mass of the particle. The force on the magnetic mass also vanishes. Note that Dirac related the magnetic ($q_m$) and electric ($q_e$) charge of an electron by his quantization rule, $q_eq_m=n\hbar/2$, where $n$ is an integer \textcolor[rgb]{0.00,0.07,1.00}{\cite{dirac}}. Should we expect an analogous relation to exit that relating the inertial mass and magnetic mass of a particle? Thereafter, the concept of a \emph{magnetic mass} has to be introduced and clarified! \textcolor[rgb]{0.00,0.07,1.00}{\cite{massm}}. Is it the mass of the magnetic monopole?

Now take the divergence of Equation  (39) and use Equation  (41) to obtain
\begin{equation}
\frac{1}{c^2}\frac{\partial^2\Lambda_D}{\partial t^2}-\nabla^2\Lambda_D=0\,.
\end{equation}
This shows that $\Lambda_D$, which has a dimension of electric field, satisfies the wave equation. It is interesting to note that the existence of $\Lambda_D$ in Equations (39) and (41) neither disturbs the continuity equation nor the wave nature of the electromagnetic field. We may associate here $\Lambda_D$ with a longitudinal wave (electroscalar) that is concomitant with Dirac's fields \textcolor[rgb]{0.00,0.07,1.00}{\cite{arb2, arb3}}. Equation (41) shows that the magnetic Dirac's field is not solenoidal as in Maxwell's theory. This reveals that Dirac's magnetic monopole are more genuine that is fundamentally associated with matter field more than with Maxwell's fields. We call the system of equations, Equations (38) - (41), the symmetrised Dirac's field equations. These are analogous to the symmetrized Maxwell's equations.  The existence of $\Lambda_D$ in the field equations above, Equations (38) - (41), will have many physical implications.

Now taking the dot product with respect to $d\vec{S}$ of Equation  (39), and integrate Equation  (41) with respect to the volume element $dV$ employing  the Stokes's and divergence theorems, and Equation  (42), we arrive at the electromotive force
$$
\int\vec{E}_D\cdot d\vec{\ell}=\int\left(\nabla^2\Lambda_D-\frac{1}{c^2}\frac{\partial^2\Lambda_D}{\partial t^2}\right)dV=0\,.
$$
Considering a complex version of Maxwell's equations, we have recently shown that one can obtain a system of equations, \emph{viz.}, \textcolor[rgb]{0.00,0.07,1.00}{\cite{arb7, arb3}}
\begin{equation}\label{1}
\vec{\nabla}\cdot\vec{E}=\frac{\rho}{\varepsilon_0}-\frac{\partial S}{\partial t}\,,
\end{equation}
\begin{equation}\label{1}
\vec{\nabla}\times\vec{E}=-\frac{\partial \vec{B}}{\partial
t}\,,
\end{equation}
\begin{equation}
\vec{\nabla}\times\vec{B}=\mu_0\vec{J}+\frac{1}{c^2}\frac{\partial
\vec{E}}{\partial t}+\vec{\nabla}S\,,
\end{equation}
and
\begin{equation}
 \vec{\nabla}\cdot\vec{B}=0\,.
\end{equation}
We argue here that a similar set-up may also exist in the framework of Dirac quaternionic theory developed above.

The symmetrised Maxwell's equations exhibiting magnetic monopole features are \textcolor[rgb]{0.00,0.07,1.00}{\cite{dirac, arb7}}
\begin{equation}\label{1}
\vec{\nabla}\cdot\vec{E}=\frac{\rho_e}{\varepsilon_0}\,,
\end{equation}
\begin{equation}\label{1}
\vec{\nabla}\times\vec{E}=-\frac{\partial \vec{B}}{\partial
t}-\vec{J}_m\,,
\end{equation}
\begin{equation}
\vec{\nabla}\times\vec{B}=\mu_0\vec{J}_e+\frac{1}{c^2}\frac{\partial
\vec{E}}{\partial t}\,,
\end{equation}
and
\begin{equation}
 \vec{\nabla}\cdot\vec{B}=\rho_m\,,
\end{equation}
where $\rho_e$ and $\rho_m$ are the electric and magnetic charge densities, and $\vec{J}_e$ and $\vec{J}_m$ are the electric and magnetic current densities, respectively.
Now comparing the system of equations, Equations (47) - (50) with Equations (38) - (41) reveals that
\begin{equation}
 \vec{\nabla}\Lambda_D=-\vec{J}_m\,,\qquad \frac{1}{c^2}\frac{\partial\Lambda_D}{\partial t}=\rho_m\,.
\end{equation}
Now take the divergence of the first equation in Equation  (51) and the partial derivative of the send equation, and use Equation  (42) to obtain
\begin{equation}
 \vec{\nabla}\cdot\vec{J}_m+\frac{\partial\rho_m}{\partial t}=0\,.
\end{equation}
Moreover, Equation  (51) discloses that
\begin{equation}
 \vec{\nabla}\times\vec{J}_m=0\,, \qquad \vec{\nabla}\rho_m+ \frac{1}{c^2}\frac{\partial\vec{J}_m}{\partial t}=0\,.
\end{equation}
It is interesting that the magnetic current and charge densities in Equations (52) and (53) satisfy  the system of generalized continuity equation, we have recently introduced \textcolor[rgb]{0.00,0.07,1.00}{\cite{hishmy}}.
 Equation (52) states that the magnetic charge is conserved. Hence, the system of equations, Equations (38) - (41) and Equations (47) - (50) are equivalent. We have shown recently that expressing Maxwell's equations in complex (quaternionic) form leads immediately to existence of magnetic charges (monopoles) \textcolor[rgb]{0.00,0.07,1.00}{\cite{arb7}}.

\subsection{\textcolor[rgb]{0.50,0.00,0.50}{Energy conservation}}

Let us now consider the system of equations, Equations (38) - (41) instead of Equations (47) - (50) and discuss the evolution of magnetic monopoles. We then would like now to see whether the scalar wave $\Lambda$ governed by Equation  (42) carries energy and momentum. To this aim we multiply (dot) Equation  (39) by $\vec{B}$ and Equation  (40) by $\vec{E}$, and subtract the two resulting equations to obtain
\begin{equation}
\frac{\partial u'}{\partial t}+\vec{\nabla}\cdot\vec{S}\,'=-\vec{J}\cdot\vec{E}\,,
\end{equation}
where
\begin{equation}
 u'=\frac{B^2}{2\mu_0}+\frac{\varepsilon_0}{2}E^2+\frac{\varepsilon_0}{2}\Lambda^2\,,\qquad \vec{S}\,'=\frac{\vec{E}\times\vec{B}-\Lambda\vec{B}}{\mu_0}\,.
\end{equation}
It is really very interesting that the new scalar wave, $\Lambda$, is a physical wave. Thus, the new system of Maxwell's equations, Equations (38) - (41), could be experimentally tested, and a possible new scalar wave accompanying the ordinary electromagnetic wave can be observed. The energy flux of this new wave points opposite to the direction of the magnetic field, and its energy density is proportional to $\Lambda^2$. In standard electromagnetic theory, magnetic monopole can be introduced in Maxwell's equations and we get a symmetrical set of equations. However, in the present formulation, magnetic monopoles are inherited and emerge naturally, in addition they give rise to scalar waves. Thus, existence of magnetic monopoles can be deduced differently. We can say that this scalar wave accompanied the magnetic charges flow.

Let us now consider a pure magnetic wave arising from magnetic charges only. In this case, $\vec{E}=0$. Hence, the energy equation in Equation  (55) yields
\begin{equation}
 \frac{\partial }{\partial t}\left(\frac{B^2}{2\mu_0}+\frac{\varepsilon_0}{2}\Lambda^2\right)+\vec{\nabla}\cdot(-\Lambda\,\vec{B})=0.
\end{equation}
We have already encountered such an equation when we dealt with longitudinal wave arising from several considerations \textcolor[rgb]{0.00,0.07,1.00}{\cite{arb2, long}}. It is also similar to Equation  (33).
It is evident that such a wave doesn't lose energy as it propagates in a conducting medium. It travels at the speed of light. It is thus the best wave that we should  use to send  information.  Moving magnetic charges produce magnetic currents that  induce electric charges (electric field). Thus, the origin of electric charge may be understood.

\section{\textcolor[rgb]{0.00,0.07,1.00}{Concluding remarks}}

Deriving Maxwell-like equations from the quaternionic Dirac equation implies that quaternionic representations are more general than vector and tensor ones. It can accommodate spin - 1/2 and spin - 1 particles in a single representation. This allows for the possibility of combining spin -1/2 particles into spin - 1, and vice versa. Note that de Broglie theory associates a wave nature with a particle, and particle nature with a wave. A complete picture would be symmetric if we associate  mass with photons. When this is done with Proca theory the gauge invariance property of Maxwell's equations is lost. In this case photons can interact with each other, and with electrons while conserving some physical quantities (like energy, spin, etc).

Owing to de Broglie theory, the photon should have inertial fields (matter wave). From  symmetric point of view, we should associate two types of fields reflecting the material nature that are analogues to the electric and magnetic field exhibiting charge nature of the particle. Our theory says that if this is done, then these material fields are governed by Maxwell-like equations as well. In this case the mass and charge of  a quantum particle can be related.
This is rather very intriguing since mass (matter) fields and charge fields obey Maxwell's equations. The matter fields governing the particle movement (motion) allows the particle to be described by  a wavefunction and other relevant quantities.

The definition we have so far made reflects this possibility. It is a kind of looking at the particle from a field point of view rather than a wavefunction. Equivalently, we treat the particle as defined by fields and not by wavefunction, as in the ordinary case of quantum mechanics. In the latter theory the interaction of the photon with the electron is done by the method of minimal substitution.
Moreover, if the quaternionic wavefunction is defined so that it incorporates electromagnetic nature of the particle, then Dirac and Maxwell's equations can be better unified within a single formulation.

The mass and current densities in our formulation reflect the material nature of the photon, but if we want these to reflect simultaneously the charge nature, mass and charge must be related. In addition, the vector and scalar  ($\vec{\psi}\,, \psi_0$) can be redefined to express electric and magnetic potentials ($\vec{A}$\,, $\varphi$) that describe electric and magnetic field of the particle. In that case the matter fields will be transformed into charge fields. Therefore, it not impossible to search for that. A faithful unified joint description of mass and charge fields of the photon will be our next battle.

The most important point in our present reformulation is the fact that the electric and magnetic  matter fields ($\vec{E}_D$\,, $\vec{B}_D$) of a moving electron obey Maxwell's equations in same way as the fields of a moving charge do. Moreover, in this formulation we describe the dynamics of a quantum particle (photon) in an equivalent electrodynamics fashion. Besides, the Lorentz boost is equivalent to the transformations of the fields as  $c\vec{\psi}\,' =c\vec{\psi}\pm\beta\vec{\alpha}\,\psi_0,\,\psi_0\,'=\psi_0\mp c\beta\vec{\alpha}\cdot\vec{\psi}\,.$

\end{document}